\documentclass[onecolumn,preprintnumbers,amsmath,amssymb,pra]{revtex4}
\usepackage{graphicx}
\usepackage{epsfig}
\usepackage{epsf}
\usepackage{amssymb}
\usepackage{dcolumn}
\usepackage{bm}
\usepackage{subfigure}
\usepackage{setspace}
\usepackage{ulem} 
\usepackage{color} 
\graphicspath{{../../../figures/}}
\def\ThisFile{\jobname}

\begin{document}
\title{Experimental Implementation of Logical Bell State Encoding}
\author{J.S. Hodges$^{1}$, P. Cappellaro$^{1}$, T.F. Havel$^{1}$, R. Martinez$^{2}$ and D.G. Cory$^{1}$}
\affiliation{(1) Massachusetts Institute of Technology, Department of Nuclear Science and
Engineering, Cambridge, MA 02139, USA \\
(2) New Mexico Highlands University, Department of Chemistry, Las Vegas, NM 87701}
\date{\today}

\newcommand{\half}{\frac{1}{2}}
\newcommand{\ket}[1]{\vert{#1}\rangle}
\newcommand{\bra}[1]{\langle{#1}\vert}
\newcommand{\ham}{\mathcal{H}}
\newcommand{\Tr}[1]{Tr\{{#1}\}}
\newcommand{\jh}[1]{\textcolor{red}{\textsf{#1}}}
\newcommand{\pc}[1]{\textcolor{blue}{\textsf{#1}}}
\newcommand{\jhC}[1]{\textcolor{red}{\textsf{#1}}} 
\newcommand{\pcC}[1]{\textcolor{blue}{\textsf{#1}}} 

\begin{abstract}
Liquid phase NMR is a general purpose test-bed for developing methods of coherent control relevant to quantum information processing.  Here we extend these studies to the coherent control of logical qubits and in particular to the unitary gates necessary to create entanglement between logical qubits.  We report an experimental implementation of a conditional logical gate between two logical qubits that are each in decoherence free subspaces that protect the quantum information from fully correlated dephasing.
\end{abstract}
\maketitle

\section{Introduction}

Experimental implementations of quantum information processors have reached sufficient complexity that it is now possible to experimentally explore the avoidance and correction of quantum errors by encoding quantum information.
These information encoding methods include active techniques, like quantum error correcting codes (QECC)\cite{PhysRevA.52.R2493,PhysRevA.54.1098,PhysRevA.54.1862,steaneQEC.1996,PhysRevLett.77.793} and dynamical decoupling \cite{BBLloyd,ViolaBB,PhysRevLett.82.2417,viola2encdyndec}, and passive techniques, such as decoherence-free subspaces (DFS) \cite{DFSZanardi,DFSGuo,DFSLidar}, noiseless subsystems (NS) \cite{PhysRevLett.84.2525}, and topological schemes \cite{Kitaev:2003qe}.  To date the ability to store information \cite{DFSPhoton,noiseless,D.Kielpinski02092001,Haffner:2005jx,langer:060502}, to perform universal quantum operations \cite{DFSevan}, and to implement simple two qubit algorithms within a variety of logical encodings \cite{mohseni:187903,LidarExp} have been demonstrated.


Entanglement, a uniquely quantum resource, enables many of the speed-ups afforded by QIP including many-body physics simulations \cite{FeynmanQC,LloydQuantSim} exponential algorithmic enhancements \cite{Shor:1994nq}, metrology \cite{cappellaro:020502} and communication \cite{BB84}.  Creation of entangled quantum states continues to drive experimental research in quantum information \cite{negrevergne:170501,lee:022330,leibfried:2005vk,haffner:2005jf} and has served as a benchmark for coherent control.
Here we combine the two concepts of logical qubits and entanglement creation to prepare a pseudo-pure version of a Bell state between logical qubits.

Control of encoded qubits must naturally respect the symmetries involved in the encoding.  In the simplest case, this is achieved by having the control Hamiltonians commute with the noise generators \cite{PhysRevLett.82.2417}.  When this is not experimentally possible or inconvenient, high fidelity control is achievable via modulation schemes that limit the encoded information's excursion out of protected subsystems to times short compared to the noise correlation time \cite{CappellaroHodgesLeakage}.

This report focuses on experimental implementations of the modulation sequences studied in  \cite{CappellaroHodgesLeakage} for creating entanglement among logical qubits, specifically creating a logical Bell state between a pair of DFS qubits immune to collective dephasing.  Furthermore, we perform our entanglement creation gate on two distinct initial states: (i) a pseudo-pure state effectively pure over the entire four qubit Hilbert space and (ii) a subsystem pseudo-pure state [see previous paper].  We also give an analysis of the quantum gate fidelities given these two input states and identify the largest errors in implementing this gate.
\section{Logical Basis Encoding}

Using the open quantum system approach, we model the total Hamiltonian of our system and environment as :
\begin{equation}
\mathcal{H} = \mathcal{H}_S \otimes \openone_{E} +\openone_{S} \otimes \mathcal{H}_E + \mathcal{H}_{SE}
\end{equation}
where $\mathcal{H}_S$ is the nuclear spin system Hamiltonian, $\mathcal{H}_E$ is the environment Hamiltonian and $\mathcal{H}_{SE}$ describes the system-environment coupling.  For this example we choose an encoding for a simple noise model: collective $\sigma_z$ noise which corresponds to random fluctuations of the local magnetic field, $B_z$.  Defining the total angular momentum of the system as $J_z = \sum_{i=1}^N \sigma_z^i$, the interaction Hamiltonian is :
\begin{equation}
	\label{Noise}
	\mathcal{H}_{SE} = \gamma J_z \otimes B_z.
\end{equation}

The potential errors that the coupling to the magnetic fields can induce belong to the interaction algebra $\mathcal{A}_z=\left\{\openone,J_z,J_z^2,..., J_z^N\right\}$ \cite{DFSevan}.  In the two spin case (N=2), the eigenspace of the noise operator $J_z$ with eigenvalue 0 is a $\mathbb{C}^2 \times \mathbb{C}^2$ decoherence free subspace and can be used to encode one qubit of information. 
This DFS is thus spanned by the basis vectors $|01\rangle$ and $|10\rangle$.  A natural encoding of a logical qubit $|\psi\rangle_L$ is given by:
\begin{equation}
	\label{encoding1}
	 \alpha |0\rangle_L + \beta |1\rangle_L \Longleftrightarrow  \alpha |01\rangle + \beta |10\rangle
\end{equation}
The logical analogs of spin operators $\sigma_x$, $\sigma_y$, $\sigma_z$, and $\openone$, which fully parametrize a single qubit, are:
\begin{equation}
	\label{EncodedOperations}
	\begin{array}{rclrcl}
		\sigma_{z}^L & \Leftrightarrow & \frac{\sigma_z^1-\sigma_z^2}{2} \ \ \ \ & \sigma_{x}^L & \Leftrightarrow & \frac{\sigma_x^1\sigma_x^2+\sigma_y^1\sigma_y^2}{2} \\
		
		\openone^{L} & \Leftrightarrow & \frac{\openone^{1,2}-\sigma_z^1\sigma_z^2}{2} \ \ \ \ & \sigma_{y}^L & \Leftrightarrow & \frac{\sigma_x^1\sigma_y^2-\sigma_y^1\sigma_x^2}{2}\\
	\end{array}
\end{equation}

Furthermore, it shall be convenient to describe conditional logic using the logical idempotent operators:
\begin{equation}
	\begin{array}{rclcl}
		E_{\pm}^L & \Leftrightarrow & \frac{\openone_L\pm \textstyle{\sigma}^L_z}{2} & = & \frac{\openone^{1,2}-\textstyle{\sigma}_z^1\textstyle{\sigma}_z^2\pm \textstyle{\sigma}_z^1\mp
		\textstyle{\sigma}_z^2}{4}
		\end{array}
\end{equation}

In the four spin case (N=4) the $J_z$ eigenspace with eigenvalue 0 is spanned by 6 basis vectors: $\ket{0011}$,$\ket{0101}$, $\ket{0110}$, $\ket{1001}$, $\ket{1010}$, and $\ket{1100}$.  Any four of these states can span a $\mathbb{C}^4 \times \mathbb{C}^4$ subspace containing two logical qubits of information.  We chose the basis $\ket{0101}$, $\ket{0110}$, $\ket{1001}$, $\ket{1010}$.  In addition to being immune to fully correlated dephasing under $J_z$, these states are also immune to pairwise collective dephasing under noise generators $j_z^{12} = \sigma_z^1 + \sigma_z^2$ and $j_z^{34} = \sigma_z^3 + \sigma_z^4$.  The protected subspace is thus a tensor product space of two qubits of the form \eqref{encoding1}.

\section{Implementing Entanglement Creation Gates on a Quantum System}

Given a fiducial state in the computational basis, applying a Hadamard gate and subsequently a CNOT gate, creates one of the four Bell states.  
The creation of a logical Bell state amounts to implementing logical versions of Hadamard and CNOT gates. The Hadamard gate on a logical qubit is specified, up to a global phase, as 
\begin{equation}
	i U_\textsf{H} = i \frac{\sigma_{x}^L + \sigma_{z}^L}{\sqrt{2}} = e^{-i \frac{\pi}{8}\sigma_{y}^L} e^{-i \frac{\pi}{2}\sigma_{x}^L} e^{i \frac{\pi}{8}\sigma_{y}^L} 
\end{equation}
The unitary operator for implementing a CNOT gate can be decomposed into a product of unitary operators of the form of single logical spin rotations and couplings of the `$ZZ$' form:
\begin{equation}
	\begin{array}{lll}
	U_{C_0 NOT} & =  & E_{+}^{1L} \sigma_x^{2L} + E_{-}^{1L} \openone^{2L} \\
	& = & e^{i \frac{\pi}{4} \openone^{1L}\openone^{2L}} e^{-i \frac{\pi}{4} \sigma_y^{2L}} e^{-i \frac{\pi}{4} (\sigma_z^{1L} + \sigma_z^{2L} ) }e^{-i \frac{\pi}{4} \sigma_z^{1L} \sigma_z^{2L}} e^{i \frac{\pi}{4} \sigma_y^{2L}}
	\end{array}
\end{equation}

By expanding each of the exponentials above using the logical Pauli operators \eqref{EncodedOperations}, many simplifications are possible.  For instance, when a logical operator consists of a sum of commuting bilinear terms, it may suffice to drop all but one of the terms in the sum and add a constant scaling factor provided this simplified unitary has the same effect as the full unitary on a state within the logical encoding.  One example is the isomorphism between a rotation of $\frac{\pi}{2}$ about $\sigma_x^{1L}$ and a $\pi$ rotation about $\sigma_x^1\sigma_x^2$ or $\sigma_y^1\sigma_y^2$:

\begin{eqnarray}
\label{FullOp}
e^{-i \theta \frac{\sigma_x^1\sigma_x^2+\sigma_y^1\sigma_y^2}{2}} & = &  e^{-i \theta \frac{\sigma_x^1\sigma_x^2}{2}} e^{-i \theta \frac{\sigma_y^1\sigma_y^2}{2}} \nonumber \\
& = & \big( \cos(\frac{\theta}{2})\openone -i \sin(\frac{\theta}{2}) \sigma_x^1\sigma_x^2 \big)\big( \cos(\frac{\theta}{2})\openone -i \sin(\frac{\theta}{2}) \sigma_y^1\sigma_y^2 \big) \nonumber \\
& = & \cos^2(\frac{\theta}{2}) \openone+ \sin^2(\frac{\theta}{2}) \sigma_z^1\sigma_z^2 -i \sin(\theta)  \frac{\sigma_x^1\sigma_x^2+\sigma_y^1\sigma_y^2}{2} \nonumber \\
& =^{L} & \cos(\theta) \openone - i \sin(\theta) (\sigma_+^1\sigma_-^2 + \sigma_-^1\sigma_+^2)
\end{eqnarray}

\begin{eqnarray}
\label{XXOp}
e^{-i \theta \sigma_x^1\sigma_x^2} & = & \cos(\theta) -i \sin(\theta)\sigma_x^1\sigma_x^2 \nonumber \\
& = & \cos(\theta)\openone -i \sin(\theta)(\sigma_+^1\sigma_+^2 + \sigma_+^1\sigma_-^2 + \sigma_-^1\sigma_+^2 + \sigma_-^1\sigma_-^2)\nonumber \\
& =^L & \cos(\theta)\openone -i \sin(\theta)( \sigma_+^1\sigma_-^2 + \sigma_-^1\sigma_+^2)
\end{eqnarray}

\begin{eqnarray}
\label{YYOp}
e^{-i \theta \sigma_y^1\sigma_y^2} & = & \cos(\theta) -i \sin(\theta)\sigma_y^1\sigma_y^2 \nonumber \\
& = & \cos(\theta)\openone -i \sin(\theta)(-\sigma_+^1\sigma_+^2 + \sigma_+^1\sigma_-^2 + \sigma_-^1\sigma_+^2 - \sigma_-^1\sigma_-^2)\nonumber \\
& =^L & \cos(\theta)\openone -i \sin(\theta)( \sigma_+^1\sigma_-^2 + \sigma_-^1\sigma_+^2)
\end{eqnarray}

Here the last lines in (\ref{FullOp}, \ref{XXOp}, \ref{YYOp}) are all equivalent.  The $=^L$ operation keeps only the terms that act within the logical subspace.  Explicitly, the logical basis states are eigenstates of the $\sigma_z^1\sigma_z^2$ operator with eigenvalue -1; likewise, the eigenvalues of $\sigma_\pm^1\sigma_\pm^2$ are all zero.
Using the following substitutions the net unitary of a Hadamard gate and a CNOT gate can be decomposed into four separate logical operations (up to a global, unobservable phase), each consisting of a $\frac{\pi}{2}$ rotation about a single bilinear term:

\begin{equation}
U_{C_0 NOT} U_H \Leftrightarrow e^{i \varphi} U_{4} U_{3} U_{2} U_{1}
\end{equation}

\begin{equation}
	\label{Eus}
	\begin{array}{rclrcl}
		U_1& = & e^{-i \frac{\pi}{4}\sigma_y^1\sigma_y^2} \ \ \ \ & U_2 & = & e^{-i \frac{\pi}{4}\sigma_y^3\sigma_y^4} \\
		U_3& = & e^{-i \frac{\pi}{4}\sigma_z^2\sigma_z^3} \ \ \ \ & U_4 & = & e^{-i \frac{\pi}{4}\sigma_x^3\sigma_y^4} \\
	\end{array}
\end{equation}

\begin{figure}[htb]
	\centering
		\includegraphics[scale=0.5]{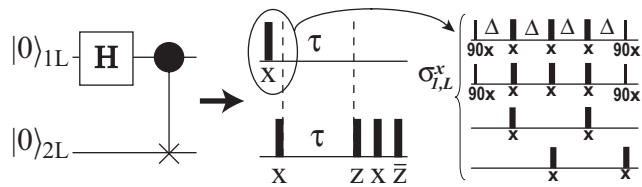}
	\label{fig:BellCircuit}
	\caption{From left to right: entangling circuit on logical qubits and corresponding logical pulses and the puulse sequence implementing the $\sigma^{1L}_{x}$ logical rotation on the physical qubits.} 
\end{figure}

The $\vec{\sigma}\cdot\vec{\sigma}$ interaction present in the internal Hamiltonian provides a means to perform $\sigma^L_{x}$ rotations, but this is efficient only when the chemical shift differences ($\vert \omega_i - \omega_0\vert$) are suppressed.
When the chemical shift differences are non-negligible a Carr-Purcell (CP) style sequence can be used to engineer a $\vec{\sigma}\cdot\vec{\sigma}$ interaction \cite{CP,CappellaroHodgesLeakage}.
By applying collective $\pi$ rotations on a pair of spins comprising a logical bit, the chemical shift terms can be averaged to zero, retaining the $ZZ$ term to first order.  Staggered $\pi$ rotations on the spins of the other logical bit are applied in a manner to refocus all operators in the Hamiltonian associated with these spins (see Fig. ~\ref{fig:BellCircuit}).

To generate a $\sigma_{x}^{1L}$ coupling, we first generate the zeroth order average Hamiltonian $\mathcal{Z}_{12}$ = $\frac{\pi}{2} J_{12} \sigma_z^1 \sigma_z^2$ \cite{AHTPhysRev,AHTHaberlen} by employing a CP-like sequence:

\begin{equation}
\label{CPSeq}
\Delta_{12} - \pi)_{x}^{1,2,3} - \Delta_{12} - \pi)_{x}^{1,2,4} - \Delta_{12} - \pi)_{x}^{1,2,3} - \Delta_{12} - \pi )_{x}^{4}
\end{equation}
(Note the notation $\theta )_\mu^k \equiv \exp[-i \frac{\theta}{2} \sum_k \sigma_\mu^k]$.)  The effective unitary propagator to zeroth order is $U^{(0)} = e^{-4 i \Delta_{12} \mathcal{Z}_{12}}$.

By applying ``collective'' $\frac{\pi}{2}$ rotations about $\hat{x}$ before the above sequence and a $\frac{\pi}{2}$ rotation about $-\hat{x}$ after, the average Hamiltonian is transformed into $\sigma_y^1\sigma_y^2$ which has the same action as a  $\sigma_{x}^{1L}$.  Similarly, if the rotation axes are separated by $\frac{\pi}{2}$ (i.e. an $\hat{y}$ phased pulse  on one of the spins in the logical pair and a $\hat{x}$ phased pulse on the other), an operator isomorphic to a $\sigma_y^{1L}$ is achieved.  The logical ``two-body'' interaction -- $\sigma^{1L}_z\sigma^{2L}_z$ -- acting on the encoded subspaces is isomorphic to $\sigma_z^1\sigma_z^3$.  We obtain this by using \eqref{CPSeq} and replacing spin 2 with spin 3.  Each of the four rotations in \eqref{Eus} can be generated in this manneryielding an overall sequence:
\begin{equation}
\frac{\pi}{2} \Big)_{\bar{x}}^{1,2} - \mathsf{Z}_{12} 
- \frac{\pi}{2} \Big)_{x}^{1,2,3,4} -  \mathsf{Z}_{34} 
- \frac{\pi}{2} \Big)_{\bar{x}}^{3,4} -  \mathsf{Z}_{23}
- \frac{\pi}{2} \Big)_{x}^{3} \frac{\pi}{2} \Big)_{y}^{4}-  \mathsf{Z}_{34}
- \frac{\pi}{2} \Big)_{\bar{x}}^{3} \frac{\pi}{2} \Big)_{\bar{y}}^{4}
\end{equation}
where $\mathsf{Z}_{jk}$ denotes the subsequence generating $\mathcal{Z}_{jk}$ and setting $\Delta_{jk} = (2J_{jk})^{-1}$.

\section{Experiment}

Our quantum system is comprised of the four $^{13}C$ spins ($I=\half$) of isotopically labeled crotonic acid (also known as trans-3-butenoic acid) \cite{CrotonicRef}.  This molecule contains nine magnetically active nuclei in total; the total spin system Hamiltonian takes the form: 
\begin{eqnarray}
	\label{Hint}
	\mathcal{H}_{internal} = \half \sum_{i \in \text{C}} (\omega_i - \omega_0^C) \sigma^i_z + \half \sum_{k \in \text{H}} (\omega_k - \omega_0^H) \sigma^i_z +\sum_{i, j \in \text{C}; i < j} \frac{\pi}{2} J_{ij}\vec{\sigma}^{i} \cdot \vec{\sigma}^{j} + \\ \nonumber
	\sum_{k,l \in \text{H}; k<l} \frac{\pi}{2} J_{kl}\vec{\sigma}^{k} \cdot \vec{\sigma}^{l} +\sum_{j,k; 	j < k} \frac{\pi}{2} J_{jk} \sigma_z^j \sigma_z^k
\end{eqnarray}
where $\omega_0^\text{C}$ ($\omega_0^\text{H}$) is the rotating frame frequency near the $^{13}C$ ( $^1$H) Larmor frequency, the $\omega_{i}$ are the chemical shifts of the 4 carbon nuclear spins,  the $\omega_{k}$ are the chemical shifts of the 5 hydrogen nuclear spins, and the 6  $J_{ij}$ (10  $J_{kl}$) are the scalar coupling constants between two carbon (two hydrogen) spins (as usual, $\hbar = 1$).  We are mainly concerned with coherently controlling the carbon subsystem of spins and seek to suppress the proton subsystem.  As the heteronuclear scalar couplings (terms $J_{j,k}$) are the only means of mixing the two subsystems, a broadband decoupling sequence modulating the proton spin system effectively removes this coupling during the experiment.  In practice, decoupling the proton spin system is equivalent to saturating the populations of the proton spins.  One potential artifact of this approach is the introduction of transient nuclear Overhauser effects (NOE) \cite{NOE1,NOE2,NOE3,NOE4}.

\subsection{Pseudo-pure states over the entire Hilbert Space}
\begin{figure}[hbt]
	\centering
		\includegraphics[scale=0.4]{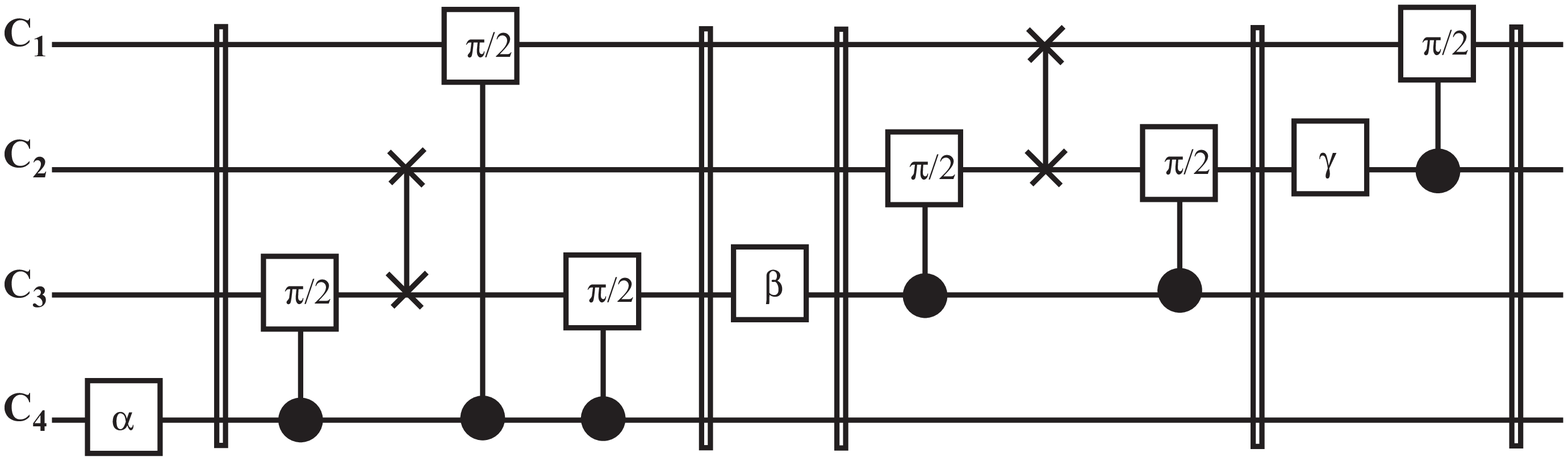}
	\caption{Circuit for the preparation of the pseudo-pure state. We represent single qubit rotations by square boxes, controlled rotations by closed circles on the controlling qubit linked to the applied rotation on the controlled qubit; swaps gate by two crosses on the swapped qubits, connected by a vertical line; non unitary operations (gradients) by double vertical bars. Notice the number of controlled operations, each requiring a time of the order of the inverse coupling strength, and \textsc{swap} gates, each requiring three times the inverse coupling strength.  The single-qubit rotations above have the values  $\alpha \approx \frac{9}{20}\pi$, $\beta \approx \frac{2}{5} \pi$, and $\gamma \approx \frac{2}{7} \pi$ which account for the scaling of the signal-to-noise compared to the equilibrium state.} 
		\label{fig:PPcircuit1}
\end{figure}

Starting with the equilibrium density matrix of the four spin system, $\rho \approx \openone/N - \epsilon(\sum_j^N \sigma_z^j)$, we prepared our system in the dual DFS ground state, $\ket{0101} = \ket{00}_L$ using spatial averaging techniques \cite{Cory1,CoryPP}.  As shown in Figure \ref{fig:PPcircuit1}, the preparation of the encoded logical state is complex and requires a preparation time of 0.1186 seconds (s) relative to the total experiment length of 0.1662 s  The $T_2$'s of the carbons are all greater than 500 milliseconds (ms) \cite{Boulant} and therefore spin-spin relaxation is unimportant over the length of the experiment.  Since pseudo-pure state preparation is a non-unitary, completely positive map, a loss of observable signal is expected.  In our implementation of pseudo-pure state preparation the signal is roughly $2/13$ that of the equilibrium state.

All qubit rotations (selective, semi-selective, or collective to all spins) were created using robust strongly-modulating pulses (SMP) \cite{softpulses,robustpulses} by maximizing the gate fidelity of  the ideal propagator to the simulated propagator.  Furthermore, as the radio frequency (RF) control fields are inhomogeneous over the sample, we maximize the effective gate fidelity, averging over a weighted distribution of RF field strengths.  Pulse lengths for this system range from 200-800 $\mu s$; the simulated gate fidelities for any individual pulse are greater than 0.99\%.  The dominant source of residual errors in a typical SMP come from two-body terms of the form $\sigma_\mu^j \sigma_\nu^k$.  After many pulses these small residual errors accumulate, but the net effect can be partially suppressed by adjusting the delay time between pulses \cite{Knill:2000fv} to optimize the overall gate fidelity or state correlation.  The required $\frac{\pi}{2}$ rotations about the logical operator axes were obtained by implementing the sequence in \eqref{CPSeq} and setting $\Delta_{jk} = \frac{1}{2J_{jk}}$.  In principle, we would like to repeat the sequence N times and scale $\Delta$ by a factor of $\frac{1}{N}$ in order to induce a rapid refocusing of the noise; however, in practice $\Delta$ is limited by the length of the semi-selective pulses.  For all of the logical rotations implemented experimentally, we use N=1.


The density matrices of the spin system after the initial pseudo-pure preparation and after the entanglement creation were reconstructed using state tomography \cite{ChuangPRSL}, which involves applying 18 readout pulses to obtain coefficients for the 256 operators comprising a complete operator basis of the four spin-$\half$ $^{13}$C nuclei.

\subsection{Subsystem pseudo-pure states}
As shown in the accompanying paper \cite{SPPS}, for mixed-state ensemble quantum information processing there are advantages to requiring the initial state to be pure only over the subsystem containing the relevant quantum information.  Here we implement the entangling operation over logical qubits using the double DFS initial state described in \cite{SPPS}.  This state (Figure \ref{Exp2}) can be prepared in half the time of the full pseudo-pure state (0.0568s) and has a smaller loss of signal (2/3 compared to 2/13).

\begin{figure}[htb]
	\begin{center}

		\includegraphics[scale=0.5]{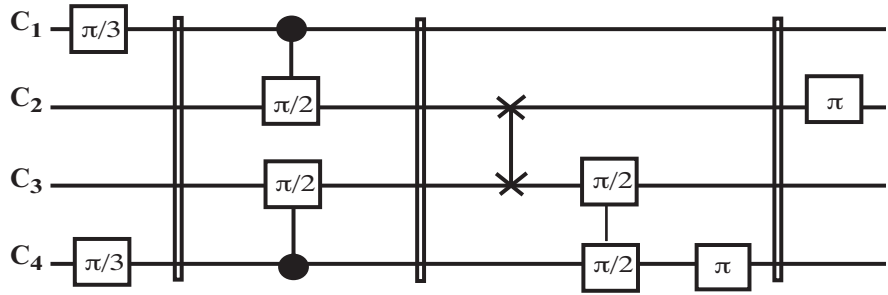}
	\caption{Circuit for the preparation of the subsystem pseudo-pure state. We use the same convention as in Figure \ref{fig:PPcircuit1}.
	}
	\label{PPcircuit2}
	\end{center}
\end{figure}

\subsection{Density Matrix Reconstruction}
\begin{figure}[htp]
  \begin{center}
    \subfigure[Subsystem Pseudo-Pure State]{\includegraphics[scale=.45]{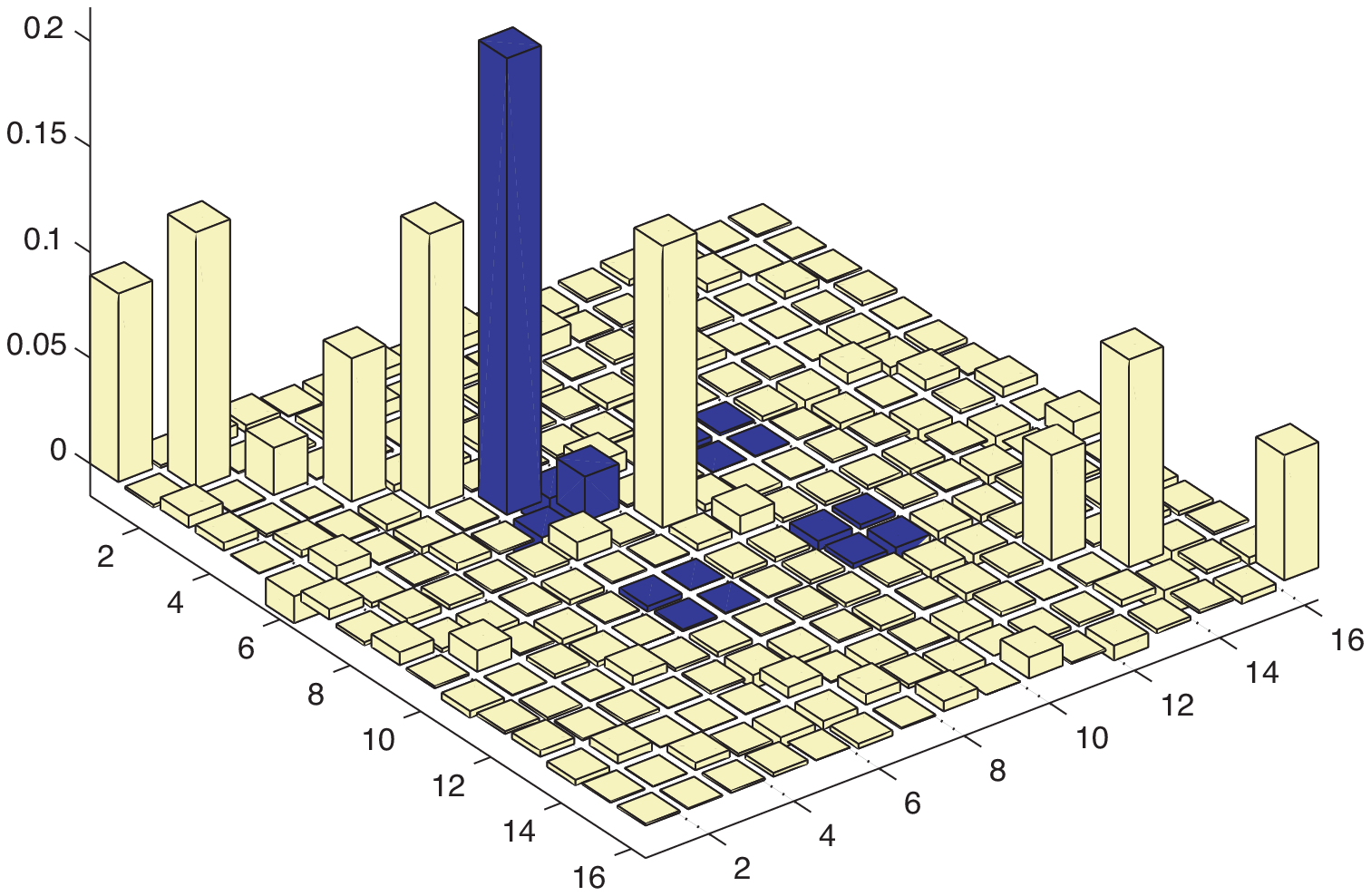}}
    \subfigure[Logical Bell State]{\includegraphics[scale=.45]{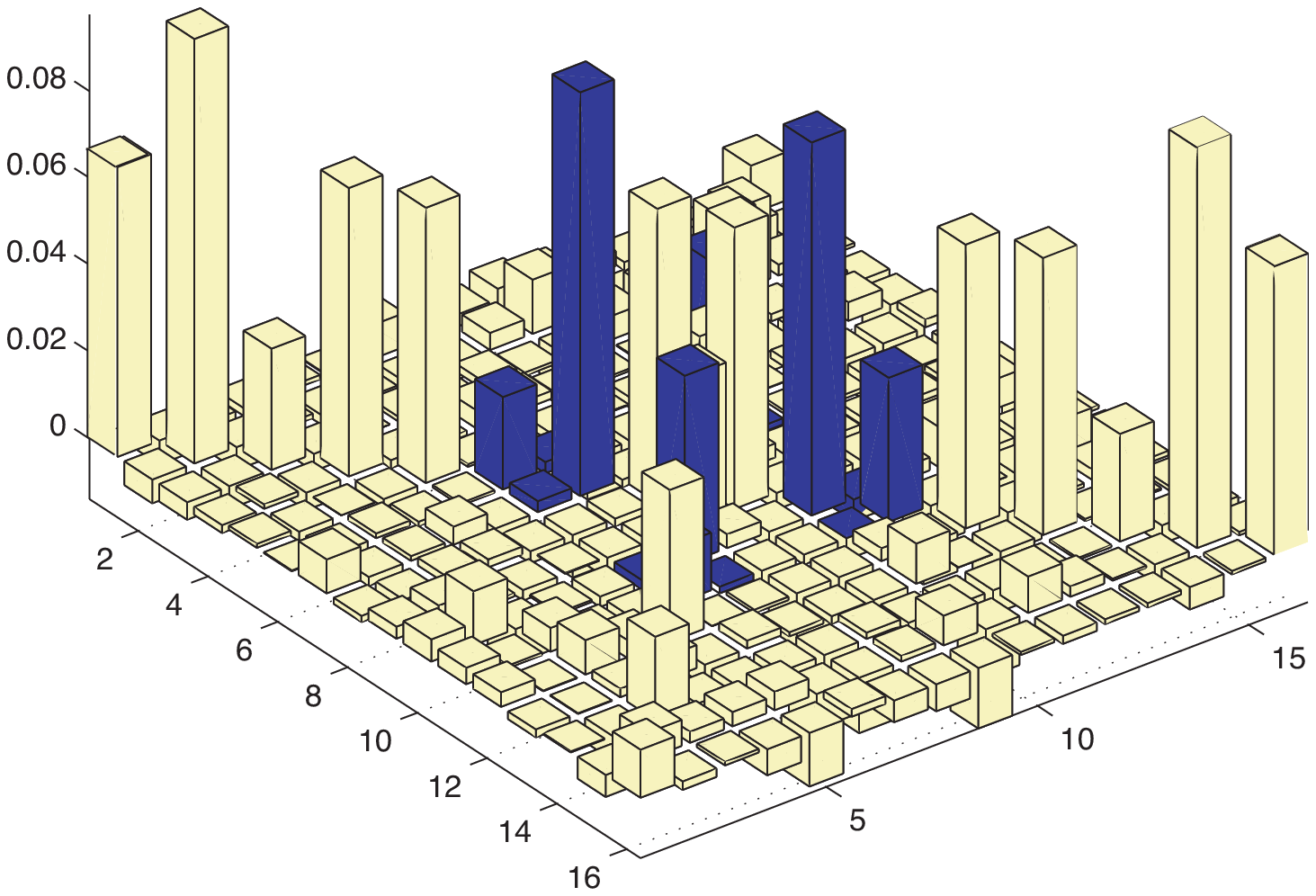}}
    \caption{Density matrices for the initial pseudo-pure state over only the logical subspace (a) and the corresponding Bell-State (b).  The darker part indicates the states in the logical subspace. In the case of the subsystem pseudo-pure stat
    es, the division of the logical subspace allows for the other areas of Hilbert space to be mixed.}
        \label{Exp2}
  \end{center}
\end{figure}

\begin{figure}[htp]
  \begin{center}
    \subfigure[Pseudo-Pure]{\includegraphics[scale=.45]{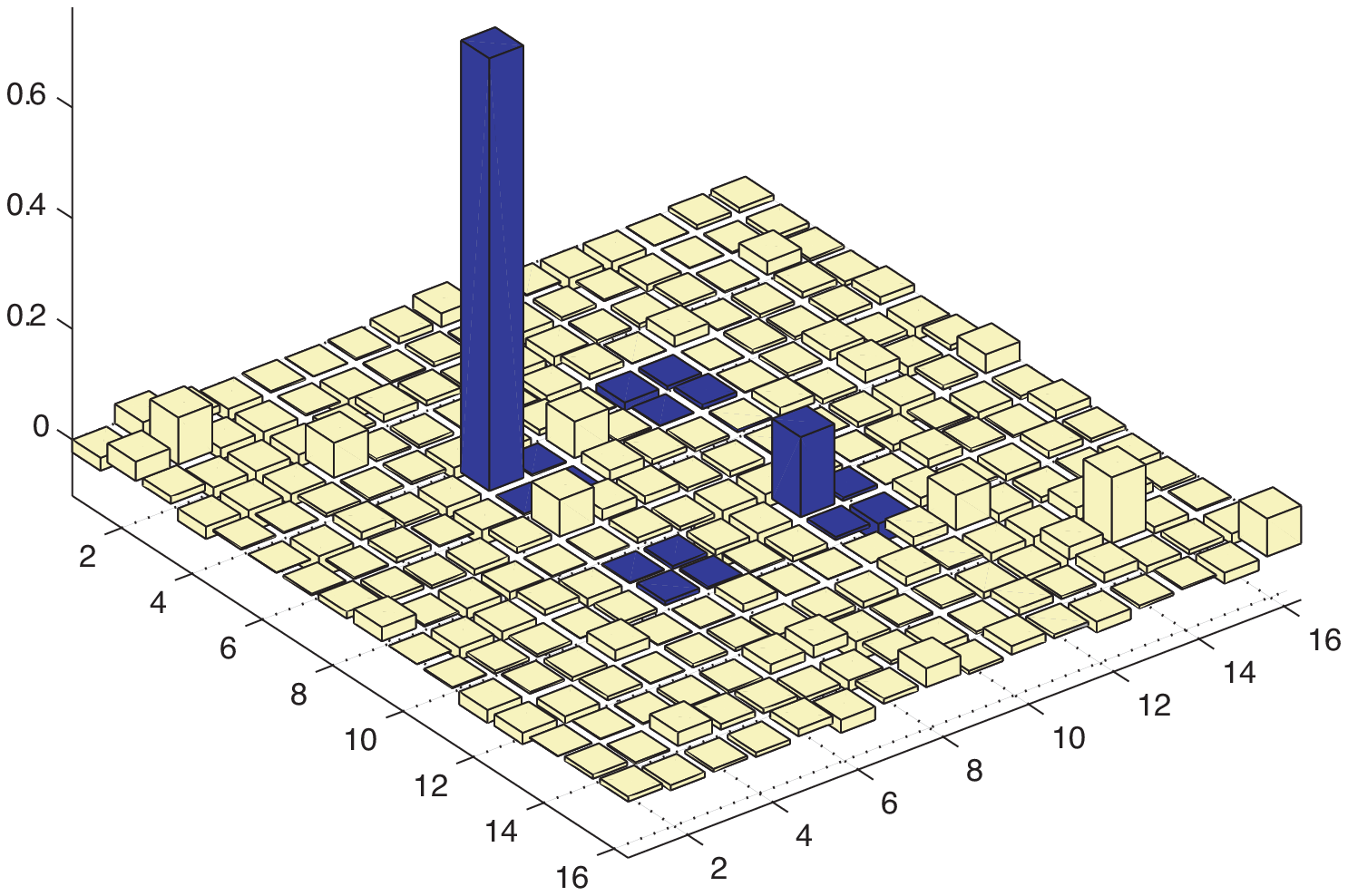}}
    \subfigure[Logical Bell State]{\includegraphics[scale=.45]{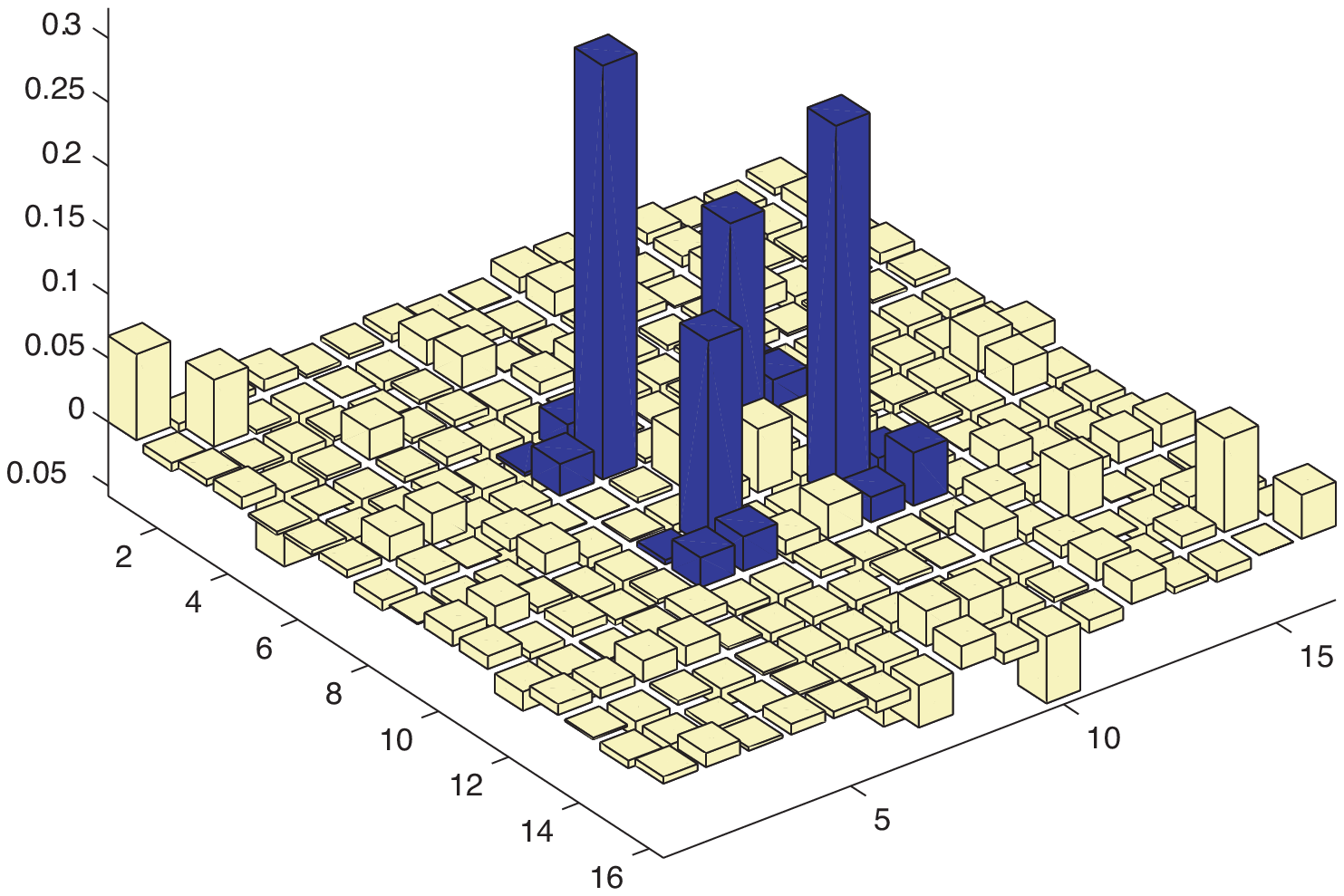}}
    \caption{Density matrices for the initial pseudo-pure state over the entire Hilbert space (a) and the corresponding logical Bell State (b).  The basis states comprising the dual DFS have been darkened.}
      \label{ExpFull}
  \end{center}
\end{figure}

The reconstructed density matrices are shown in Figures \ref{Exp2} and \ref{ExpFull}, where the vertical axis shows the normalized amplitude and the horizontal axes label the basis states in the computational basis (i.e. $\ket{0000}$,$\ket{0001}$,...).  
The effects of decoherence can be qualitatively seen in the final state as an attenuation of the off-diagonal terms of the Bell state: $\frac{\ket{00}_L - \ket{11}_L}{2}$.

\section{Discussion}
\subsection{Metrics of Control}

The quantum process associated with the encoded entangling operation can be specified by the general map:
\begin{equation}
\mathcal{E}(\rho) = \sum_\mu A_\mu \rho A_\mu^\dag
\end{equation}
where $A_\mu$ are Kraus operators describing the experimental implementation of the encoded operation, and $\rho$ is defined over the entire Hilbert space.  If the process contains no error only a single Kraus operator defines the operation: $A_0= U$.  The correlation of two quantum states is defined as:
\begin{equation}
\label{corrdef}
	C = \frac{\Tr{\rho_{th} \rho_{exp}}}{\sqrt{\Tr{\rho_{th}^2}\Tr{\rho_{in}^2}}}
\end{equation}
which defines the relative closeness of states in Hilbert space, taking into account attenuation due to decoherent or incoherent processes.  Here $\rho_{th}=U\rho_{in}U^\dag$, $\rho_{exp} = \mathcal{E}(\rho_{in})$, and $\rho_{in}$ define the theoretical, experimental and input states respectively.  The closeness of the theoretical state to the experimental state measures how well we have implemented the logical entangling operation for a particular input state.  We reserve the symbol $C^\prime$ for designating the correlation that does not take into account the attenuation of the input state; in this measure $\rho_{in}$ is replaced by $\rho_{exp}$. 

When the relevant quantum information resides only in a subspace, we can also measure the correlation inside this subspace by applying the projectors $P_\textsf{L}$ spanning the subspace to the state \cite{SPPS}:
\begin{equation}
	C_\textsf{LL} (\rho) =\frac{\Tr{P_\textsf{L}\rho_{th}P_\textsf{L} \sum_\mu A_\mu
	P_\textsf{L} \rho_{in} P_\textsf{L} A_\mu^\dag }}
		{\sqrt{\Tr{(P_L\rho_{in}P_L)^2} \Tr{(P_L\rho_{th}P_L)^2}}} 
\end{equation}
This measure is important when creating subsystem pseudo-pure states, where nonzero contributions to the density operator outside the logical subspace can degrade $C$.  If the input state of this process is a full pseudo-pure state (FPPS), then $\rho_{in} = P_\textsf{L} \rho_{in} P_\textsf{L}$.  Furthermore, the result of a logical operation acting on a FPPS adheres to the relation: $\rho_{th} = P_\textsf{L} \rho_{th} P_\textsf{L}$.  Under these conditions the correlation $C_\textsf{LL}$ with the logical ideal state is just the total correlation $C$.

Finally, we can consider the correlation of the state after (i) an ideal decoding operation, $U_\text{Dec}$ which transforms the subspace structure of the Hilbert space back to a tensor product structure and (ii) a partial trace over the ancilla.  This examines how the entanglement created under two logical qubits transfers back to physical qubits.  For the encoding chosen in this paper $U_\text{Dec}$ is:
\begin{eqnarray}
U_\text{Dec}  & = &C_0^1NOT^2 \cdot C_0^3NOT^4 \nonumber \\
& = & \big( E_+^1\sigma_x^2 + E_-^1\openone^2 \big) \big( E_+^3\sigma_x^4 + E_-^3\openone^4 \big)
\end{eqnarray}
  $C_\text{Dec}$ has the same form as \eqref{corrdef} with  $\rho_{th}$, $\rho_{exp}$, and $\rho_{in}$ replaced by $\rho_{th}^d$, $\rho_{exp}^d$, and $\rho_{in}^d$ where $\rho_{[\;]}^d = tr_a \{U_\text{Dec} \rho_{[\;]} U^\dagger_\text{Dec} \}$ and $a$ is the subsystem spanned by the ancilla qubits.  These values are displayed in Table \ref{tab:corr}.

\begin{table}
\begin{tabular}{lcccccccc}
&   &   &  & & & &\vspace*{-3mm}    \\
${\sf Quantum\ state\;\ \ }$ & $\ C_{\text{Sim}}\ \ $
                      & $\ C^\prime_{\text{Exp}}\ \ $
                      & $\ C_{\text{Exp}}\ \ $
                      & $C^\prime_{\textsf{LL}}$
                      & $C_{\textsf{LL}}$
                      & $C^\prime_{\text{Dec}}$
                      & $C_{\text{Dec}}$
                       \\ \hline\hline
                  &   & & & &   &   &\vspace*{-3.5mm}   \\
 Full pseudo-pure (FPPS) & 0.950 & --- & 0.884  & --- & 0.981 & --- &0.986  \\ 
 Full pseudo-pure Bell state (FPPBS)  & 0.747 & 0.685 & 0.531 & 0.904 & 0.590  & 0.852 &  0.641 \\  \hline
 Subsystem pseudo-pure (SPPS) & 0.984 & --- & 0.965  & --- &0.988 & --- &  0.992 \\ 
 Subsystem pseudo-pure Bell state (SPPBS) & 0.919 & 0.836 & 0.725 & 0.849 &0.608 & 0.974 & 0.823\\ 
\end{tabular}

\vspace*{1mm}
\caption{Experimental and simulated data for the implementation
of encoded Bell state propagator. Experimental errors of $\approx 4\%$ can be attributed to systematic errors in the fitting algorithm used to reconstruct the density matrix from NMR spectral data.  The correlations of the ideal logical Bell state are reported with respect to the following states: a simulation of the spin system with  distribution of RF powers (${\cal C}_{Sim}$), the experimental tomography reconstruction of the state in the entire Hilbert space (${\cal C}_{Exp}$), the the experimental reconstruction projected onto the logical subspace (${\cal C}_\textsf{LL}$).  ${\cal C}_\text{Dec}$ compares the ideal state and the reconstructed state assuming ideal decoding out of the logical subsystem and tracing over the ancillae.  Both the figures reported for the Full pseudo-pure and Subsystem pseduo-pure Bell states reflect attenuation due to relaxation processes. Primes ($\prime$) indicate unattenuated correlations.}
\label{tab:corr}
\end{table}

Since the same pulse sequence implementing the entanglement creation was used for both initial states, differences in the correlation are due the unwanted evolution of initial errors in state preparation.  These  differences reveal several key features of the experimental implementation.  In comparing the correlation of the simulated resutls to the experiment, we see the smallest deviations in the case of the subsystem pseudo-pure state (SPPS), attributing this to the benefits of a shorter preparation sequence.  We also note that $T_1$ and $T_2$ processes, the NOE and RF Transients are not contained in the simulation model.  It follows that the longer and more complex preparation sequence for the full pseudo-pure state (FPPS) should result in a larger deviation of the correlation.  For the simulation of the full pseudo-pure Bell state (FPPBS) and the subsystem pseudo-pure Bell state (SPPBS), a reduction from unity is due solely to incoherent (RF inhomogenities) and coherent errors; the experimental correlations are further affected by the processes stated above, as well as $T_2$ processes.  

In comparing the experimental logical subspace correlation ($C_\textsf{LL}$) to $C$, we can better understand the source of error.  For instance, for the SPPS, FPPS, and FPPBS, $C_\textsf{LL}$ is greater than $C_{exp}$.  This difference represents errors in $C_{exp}$ due to unwanted, nonzero contributions to the density operator \textit{outside} the logical subspace since $C = \sum_\textsf{jk} \alpha_\textsf{jk} C_\textsf{jk}$ where $C_\textsf{jk}$ represent correlations for blocks of the density operator and $\alpha$ is a weighting factor \cite{SPPS}.  However, in the case of the SPPBS, $C_\textsf{LL}$ is less than $C_{exp}$.  Here $\alpha_\textsf{LL}$ = 0.40, $\alpha_\textsf{RR}$ = 0.58, and $C_\textsf{RR}$ = 0.822.  Thus by removing the $C_\textsf{RR}$ contribution from $C_{exp}$ we get a more complete view of the control over the logical qubits.  Consequently for this state we controlled the information in the non-logical subspace better than the information inside the subspace of interest.  We believe the source of this difference to be $T_2$ decoherence, as the majority of \textit{information} in the $\mathcal{R}$ subspace exists as population terms ($\sigma_z$, $\sigma_z\sigma_z$, etc).  In the case of fully correlated noise, the characteristic decoherence time for any state with $J_z =0$, $T_2^0$, should approach the characteristic relaxation time of the population terms ($T_1$).  For this implementation, the $T_1$ times for any individual spin are greater than 2 seconds, whereas the shortest $T_2^0$ is 1.1 seconds which is consistent with our state tomography measurements. 

Lastly, we comment on the differences between the $C_\textsf{LL}$ and $C_\text{Dec}$.  Both of these metrics assess the closeness of two states in a $\mathbb{C}^2 \otimes \mathbb{C}^2$ space, but in a different manner.  In the absence of relaxation, the decoded correlation assesses how well the implemented entangling operation respects the subspace structure of the system, namely $U_{ent} = U_{ent_L} \oplus U_{ent_R}$.  
Under the decoding, this operation should only create entanglement among two physical qubits and thus the ancillae can be traced over.  To the extent that $U_{ent}$ does not respect the subspace structure, the decoding will in general introduce entanglement between the informational qubits and the ancillae, thus resulting in a loss of purity after a partial trace.  In the case of the FPPBS, $C_\text{Dec}$ is larger than $C_\textsf{LL}$ due to a loss of purity of the state; however, $C^\prime_\text{Dec}$ is smaller than $C^\prime_\textsf{LL}$.  This shows that the sensitivity to coherent unitary errors and incoherent/decoherent errors of the two metrics differs slightly.  In the case of the SPPBS, the correlation and unattenuated correlation of the decoded metric are greater than their respect logical projection metrics.  This can be attributed to the contributions to $\rho^\prime_{exp}$ from state preparation. (Recall that for a SPPS, the decoded state is not pure on the information carrying qubits, even in the ideal case.)  These contributions are less sensitive to errors during the quantum operation and thus augment the value of the decoded correlation.

\begin{table}
\begin{tabular}{lcc}
&  & \vspace*{-3mm} \\
\textsf{Quantum State} & $tr(\rho_{exp}^2)$ & $tr(\rho_{sim}^2)$ \\
\hline \hline
Pseudo-pure & 0.781 & 0.812 \\
Logical Bell State & 0.470 & 0.556 \\ \hline
\end{tabular}
\vspace*{1mm}
\caption{
The above values correspond to the purity of the initial and final states for the experimentally derived density operator and that obtained by simulating the four carbon system with RF inhomogeneity.  Note that the purity of the initial state is not unity, as small errors to the eigenspectrum of $\rho$ are quadratically amplified by the trace.  The loss of purity from the initial to the final state for the simulated gates does not take into account decoherence; thus, any loss of purity must arise from the incoherence due to RF inhomogeneity.}
\label{tab:purity}
\end{table}

\section{Conclusion}

Using a two-physical qubit which protects against collective dephasing, we have shown how to implement quantum gates between logical qubits using effective Hamiltonians.  It is important to stress that when selecting a protection scheme against decoherence the ability to encode a physical qubit into a logical qubit with high fidelity is not sufficient for computation.  The structure of the external and natural Hamiltonian plays an important role in the control of logical qubits, as the operators needed to implement gates may not be present and generating them may drive the information out of the subsystem.  For large systems with significant symmetry (like quantum dots under the exchange interaction \cite{DiVincenzoExchange}) or exceedingly small systems \cite{DFSevan}, the structure of the natural Hamiltonian can provide the logical operations in itself.  However, for systems of intermediate size (most relevant to the present implementations of quantum information processors) implementing quantum gates among logical qubits requires both a precise knowledge of the system Hamiltonian and a set of control parameters large enough to ensure no leakage from the protected subsystem or subspace.  For example, if our four-qubit system were composed of two protons and two carbons, each individual species could be modulated separately, thus doubling the number of parameters in the control Hamiltonian and limiting the leakge of the information from the subspace.

Finally, our selection of logical qubits comprised of only two physical qubits limits our logical operations for single qubit and two-qubit interactions to only ``two-body" operators.  If instead we were to attempt a repetition of the experiment where the logical qubits were encoded under the three qubit noiseless subsystem \cite{noiseless}, the single qubit and two-qubit rotations would involve ``three-body" operators \cite{PhysRevA.61.012302} -- quite unlikely to be found in a natural Hamiltonian.  In such a scenario, the ability to implement logical operations would necessarily need to come from a modulation sequence, appropriately chosen to avoid leakage from the subsystem.  

\section{Acknowledgements}This work was supported in part by the National Security Agency (NSA) under Army Research Office (ARO)
contract numbers  W911NF-05-1-0469 and DAAD19-01-1-0519, by the Air Force Office of Scientific Research, and by the Quantum Technologies Group of the Cambridge-MIT Institute.  We also thank N. Boulant for useful discussions.
\bibliography{EncOpsNoUrl.bib}

\end{document}